# Novel procedure to prepare cadmium stannate films using spray pyrolysis technique for solar cell applications

V. Krishna Kumar,[1] K. Ramamurthi[1*] and E. Elangovan[2]

[1]Department of Physics, Bharathidasan University, Tiruchirappalli - 620 024, India

[2]Materials Science Department/CENIMAT, Faculty of Science and Technology, New University of Lisbon, Campus da Caparica, 2829-516 Caparica, Portugal.

Thin films of cadmium stannate was prepared using low cost cadmium acetate and tin (II) chloride precursors by spray pyrolysis technique at three different substrate temperatures of 400, 450 and 500$^0$ C. A novel procedure of simultaneously forming additional layer, introduced for the first time in this work, on the already coated cadmium stannate film reduced the sheet resistance from 160 Ω/sq to 15 Ω/sq. Further, it is identified that the formation of additional layer does not affect the structural and optical properties of the cadmium stannate films, but improves the electrical property; thus the formation of additional layer seems to be an effective alternate for annealing the films at high temperature in the presence of Ar, CdS, Ar/CdS mixture, hydrogen or nitrogen to improve the structural, electrical and optical properties of the cadmium stannate films as has been reported in the literature. The maximum optical transmittance value of the prepared cadmium stannate film is about 99.8 % and the optical band gap energy value is about 2.9 eV.

PACS number(s): 81.15.Rs, 61.10.Nz, 73.61.Ph, 78.66.Jg

# I. INRODUCTION

Cadmium Stannate ($Cd_2SnO_4$) or Cadmium Tin Oxide (CTO) is an n- type semiconductor.[1] Thin films of cadmium stannate have great technological interest due to their high quality electrical and optical properties superior to the conventional transparent conducting oxide materials.[2,3] These films have wide applications in photogalvanic cells,[4] liquid crystal displays,[5] heat mirrors[6] and transparent electrode.[7] $Cd_2SnO_4$ thin films have several significant advantages in the CdTe/CdS based solar cells. Comparing with $SnO_2$, cadmium stannate films have better adhesion in the CdTe/CdS solar cells.[8] The reduced film thickness of $Cd_2SnO_4$ films in the solar cell devices reduces the number of interconnects and reduces manufacturing costs.[9]

Preparation of cadmium stannate material was first reported by Smith[10] employing the procedure of solid-state reactions whereas the preparation of cadmium stannate thin film was first reported by Nozik[1] by rf sputtering. Subsequently various deposition methods of rf sputtering,[1,2,6,7,11-17] DC reactive sputtering,[18-21] DC magnetron plasmatron sputtering,[5] reactive magnetron sputtering[22] and atom beam sputtering[23] were employed to prepare cadmium stannate films. Sputtering atmosphere used was Ar[1,18] or Ar-$O_2$ mixture[1,13,14,19] in the pressure range of $2.0 \times 10^{-2} - 8.0 \times 10^{-2}$ Torr and the substrate temperature was varied from room temperature to $400^0$ C. Prepared films were then annealed[1,3,5,7,11-15,19,24,25] in the presence of Ar,[5,19,25] Ar/CdS mixture,[7,12,13,23] hydrogen gas,[1,23] nitrogen,[11,25] CdS[2] or in vacuum[14,15,24,25] ($10^{-5}$ Torr) to increase the free electron concentration[12,15] and optical transmittance.[5,14,16,25] Further the process of annealing crystallizes the amorphous films obtained in almost all the above deposition techniques.

Employing spray pyrolysis technique Haacke et al.[26] prepared cadmium stannate films from the aqueous solutions of $CdCl_2$ and $SnCl_4$ at substrate temperatures greater than $800^0$ C. Mohammad and Abdul Ghafor[27,28] and Agnihotri[24] sprayed the precursor solution containing the mixture of $CdCl_2.5H_2O$ solution in ethyl alcohol and $SnCl_4.3H_2O$ solution in acetic acid onto the heated substrates of glass or quartz under normal atmospheric conditions; this produced amorphous films of cadmium stannate. Then the prepared films were heated to $300^0$ C under vacuum of $10^{-5}$ Torr to reduce the effect of oxygen traps created during film preparation. Thus a survey of literature reveals that cadmium stannate films prepared by various techniques were annealed at elevated temperatures of about 600 -$800^0$ C in the atmosphere of some gases or at high vacuum to improve their electrical, optical and structural properties.[2,5,12,14-16,25]

The present work was carried out with an aim to identify the possibility of preparing cadmium stannate thin films from a set of new and low-cost precursors using home made spray pyrolysis technique. Hence an attempt was made to prepare cadmium stannate thin films from various starting materials of tin and cadmium compounds. Different molar ratio and weight ratio of various tin and cadmium compound precursors were dissolved in different solvents and were employed to prepare these thin films at different substrate temperatures. Finally it was identified that a combination of tin (II) chloride and cadmium acetate precursors produces transparent and conducting single phase cadmium stannate films.

Thus in this article we report for the first time the preparation of cadmium stannate films by spray pyrolysis technique using low cost precursors without post heat treatment. We also report for the first time a novel procedure of simultaneously forming an additional layer to improve the electrical properties of cadmium stannate films. In this way formation of additional layer seems to be an effective alternate method for post-annealing these films.[1,3,5,7,11-15,19,24,25]

## II. EXPERIMENT

### A. Experimental setup

Cadmium stannate thin films were prepared by the homemade spray pyrolysis setup.[29] The parameters such as distance between the spray nozzle and the substrate (35cm), and spray angle (about $45^0$) were kept constant. Films were coated for various spray time and spray intervals. The carrier gas flow rate was maintained at 6 l/min at a pressure of $6.5 \times 10^4$ $Nm^{-2}$.

### B. Film Preparation

Spray aqueous solution was prepared by mixing $Cd(CH_3COO)_2 \cdot 2H_2O$ and $SnCl_2 \cdot 2H_2O$ in the molecular weight ratios of 2:1, 4:1 and 6:1 respectively. A few ml of concentrated hydrochloric acid was added to get clear solution. The resultant solution was sprayed on the preheated glass substrates at various substrate temperatures of 400, 450 and $500^0C$ to prepare cadmium stannate thin films. Film thickness was estimated employing gravimetric method. Computer controlled Philips X Pert PRO X-ray diffraction system was employed to record X-ray diffraction of these films using $CuK_\alpha$ radiation of wavelength 1.5405 Å. Sheet resistance of the films was measured using four-probe experimental setup and the optical transmittance spectrum was recorded using Shimazdu, UV-1601, UV-Visible spectrometer in the range of 300 – 1100nm.

### C. Formation of additional layer

Decomposition of the sprayed source materials was observed only when the substrate temperature was above $350^0C$ whereas, undecomposed opaque coatings resulted when the substrate temperature was below $300^0$ C. Hence in the process of formation of additional layer, films were coated by spraying the precursor solution on the heated substrates at 400, 450 or $500^0$ C by retaining a few ml of the spray solution. Then the substrate heater was switched off immediately. When the substrate temperature was decreased below $300^0$ C the retained remaining solution was sprayed on the films formed already. Thus an additional layer of white opaque coating was formed on the surface of the coated films. Then the temperature of the substrate was allowed to cool to the room temperature. This additional layer was removed easily by two ways with out affecting the already coated cadmium stannate films: (i) The coated films were kept in the moisture rich atmosphere at about $16^0$ C. The opaque coating on absorbing the moisture automatically started to remove from the surface of the coated films. Then the substrate was washed with a small quantity of methanol. (ii) The additional layer was simply cleaned by using methanol without placing it in the moisture rich atmosphere. The films prepared by following the process of formation of additional layer were found to retain their optical and structural properties with improved electrical properties when compared to the films coated without additional layer.

## III. RESULTS

Out of many experiments carried out by us a few results are presented in Table I along with the spraying conditions and the electrical properties of cadmium stannate films. Even though the films prepared at $450^0$ C are amorphous, they are also presented in Table I for the purpose of comparison.

TABLE I. Preparation parameters and electrical properties of cadmium stannate thin films.

| Film | Spraying Condition | | | Substrate Temperature °C | Thickness (μ m) | Resistivity (Ω cm) | Sheet resistance Ω/sq. | Conductivity (Ω$^{-1}$ cm$^{-1}$) | Additional layer |
|------|--------|--------|--------|------|------|------|------|---------|-----|
|      | W. Ratio | Spray Interval (Sec) | Spray Time (Sec) | | | | | | |
| A | 2:1 | 60 | 1   | 450 | 0.2  | 5800 | 29K  | 1.72    | -- |
| B | 4:1 | 60 | 1   | 450 | 0.18 | 198  | 1100 | 50.50   | -- |
| C | 4:1 | 30 | 0.5 | 450 | 0.16 | 128  | 800  | 78.13   | -- |
| D | 6:1 | 30 | 2   | 450 | 0.3  | 60   | 200  | 166.67  | -- |
| E | 6:1 | 30 | 2   | 500 | 0.34 | 69   | 230  | 144.93  | -- |
| F | 6:1 | 30 | 3   | 500 | 0.21 | 33.6 | 160  | 297.62  | -- |
| G | 6:1 | 30 | 4   | 500 | 0.22 | 37.4 | 170  | 267.38  | -- |
| H | 6:1 | 30 | 3   | 500 | 0.28 | 6.6  | 29   | 1515.15 | Yes |
| I | 6:1 | 30 | 3   | 500 | 0.4  | 6    | 15   | 1666.67 | Yes |

### A. Structural Properties

X-ray diffractogram (XRD) of films coated at the substrate temperature of 450 and $500^0$ C are shown in Fig.1. The literature survey shows that as deposited cadmium stannate films are amorphous in many experiments. By annealing these films at different temperatures in the atmosphere of various gases or mixture of gases polycrystalline films were obtained. Wu et al.[2] observed that as deposited films by RF sputtering are amorphous and crystallization of the prepared cadmium stannate films takes place during the annealing treatment at a temperature about $580^0$ C and the intensity of the crystalline peaks increases with annealing temperature. Miyata et al.[20] observed that as the rate of deposition changes the structure of cadmium stannate film also changes. They reported that XRD patterns showed the (001) and (130) peaks of $Cd_2SnO_4$ with the deposition rate of 40 Å min$^{-1}$ and the (200) peak of $CdSnO_3$ with the deposition rate of 20Å min$^{-1}$. XRD recorded for the films coated in this work confirms the formation of $Cd_2SnO_4$ single phase thin films when compared with the corresponding XRD of the JCPDS data.[30] Films prepared at a substrate temperature of 450$^0$ C are amorphous in nature [see Fig. 1(a)] whereas the films prepared at 500$^0$ C have preferred (111) orientation of orthorhombic structure [see Fig. 1(b)]. Further one can observe from the results of XRD that the formation of additional layer [see Fig. 1 (c)] does not affect the structural properties when compared with that of the films prepared without formation of additional layer [see Fig. 1(b)].

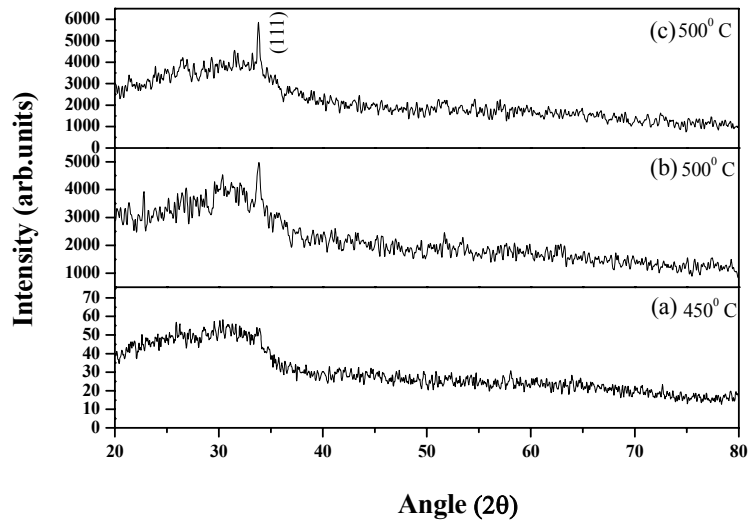

FIG. 1. XRD of the cadmium stannate films

## B. Electrical Properties

Electrical properties of these films depend on the molecular weight ratio of cadmium acetate and tin (II) chloride in the starting solution, spray time, formation of additional layer, thickness of the film and the substrate temperature.

### 1. Influence of precursors

Variation of the electrical properties of these films was found to depend on the concentration of cadmium acetate and tin chloride in the starting solution. A few drops of concentrated hydrochloric acid was carefully added to the starting aqueous solution containing cadmium acetate and tin (II) chloride just to obtain clear solution. It is observed in this work that, films prepared from the solution containing cadmium acetate and tin (II) chloride in the molecular weight ratio of 2:1 respectively have very high sheet resistance in the range of 2.5 –25 K $\Omega$/sq depending on the film thickness. Further sheet resistance is found to be in the range of 500 - 300 $\Omega$/sq for 4:1 weight ratio and 700-150 $\Omega$/sq for 6:1 ratio. Thus the results of this work reveal that as the concentration of cadmium acetate increases in the starting solution the sheet resistance decreases. The films prepared with 6:1 ratio with the spray time of 3 seconds and the spray interval of 30 seconds have good optical and electrical properties (Table I).

### 2. Formation of the additional layer

Films prepared without formation of additional layer have a minimum sheet resistance of about 160 $\Omega$/sq. But the films prepared with the formation of the additional layer have a minimum sheet resistance of about 15 $\Omega$/sq Agnihotri et al.[24] reported that excess amount of oxygen introduced into the films during the film preparation acts as electron trap and therefore these films were annealed to reduce the sheet resistance. But in the present investigation the formation of the additional layer at the temperature of $300^0$ C prevents the direct contact of the film with the atmosphere and hence the absorption of the excess amount of oxygen from atmosphere. Sheet resistance and thickness of cadmium stannate films prepared by earlier workers by employing various methods are compared with the corresponding values of this work in Table II.

TABLE II. Comparison of the sheet resistance.

| S. No | Method | Film Thickness ($\mu$m) | Sheet resistance ($\Omega$/sq) | Reference |
|---|---|---|---|---|
| 1. | rf sputtering | 3.3 | $2.3^a$ | 1 |
| 2. | DC sputtering | 0.3 | 40 | 18 |
| 3. | DC sputtering | 0.22 | 180 | 18 |
| 4. | DC sputtering | 4.4 | $14^a$ | 19 |
| 5. | rf sputtering | 0.4 | 14 | 30 |
| 6. | rf sputtering | 0.51 | $2.5^a$ | 2 |
| 7. | Spray Pyrolysis | - | 80-100 | 26 |
| 8. | Spray Pyrolysis | 0.4 | 15 | Present work |

$^a$Post-annealed films.

### C. Optical Properties

Optical transmittance spectrum of cadmium stannate films prepared at different spray conditions are presented in Fig. 2. These films have the transmittance in the range of 91.7-99.8% in the wavelength range of 665nm – 875nm. The maximum transmittance of 99.8% is obtained in the wavelength range of 770-795 nm for the amorphous film (C) of thickness of 0.16 µm prepared at the substrate temperature of $450^0$ C. The other films G,

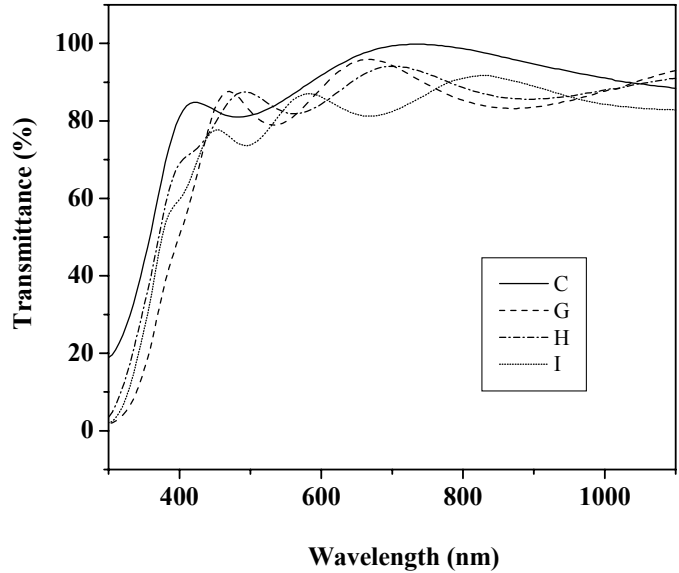

FIG.2. Optical transmittance spectrum

H and I are prepared at substrate temperature of $500^0$ C and the films H and I are prepared with the formation of additional layer (see Table I). It is evident that the formation of additional layer does not affect the optical transmittance of the films (see Fig.2). The absorption co-efficient $\alpha$ is calculated from the transmittance spectrum using the relation,

$$\alpha = -[\log_e(1/T)]/t,$$

where, T is the transmittance and t is the thickness of the film. Variation of optical absorption coefficients with photon energy is given in Fig. 3 which shows that absorption co-efficient exponentially decreases as photon energy decreases.

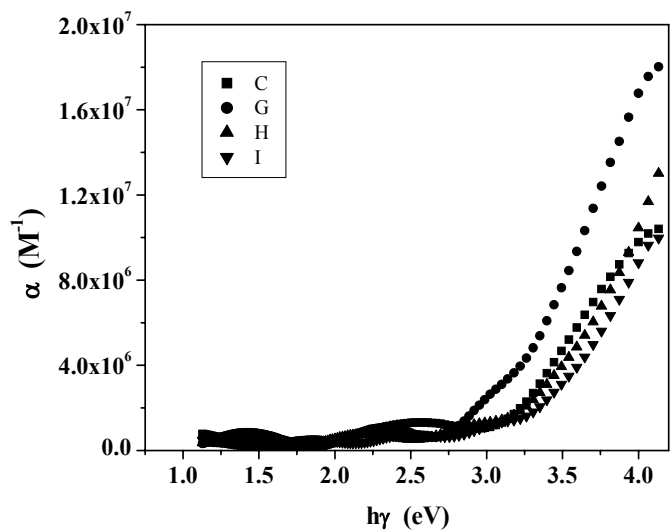

FIG. 3. Variation of optical absorption coefficient

A plot drawn between $(\alpha h\gamma)^{1/2}$ and $h\gamma$ is shown in Fig.4. Optical band gap energy for the film prepared at the substrate temperature of $500^0$ C is 2.9 eV and all the other films reported in this work also have optical band gap around 2.9 eV.

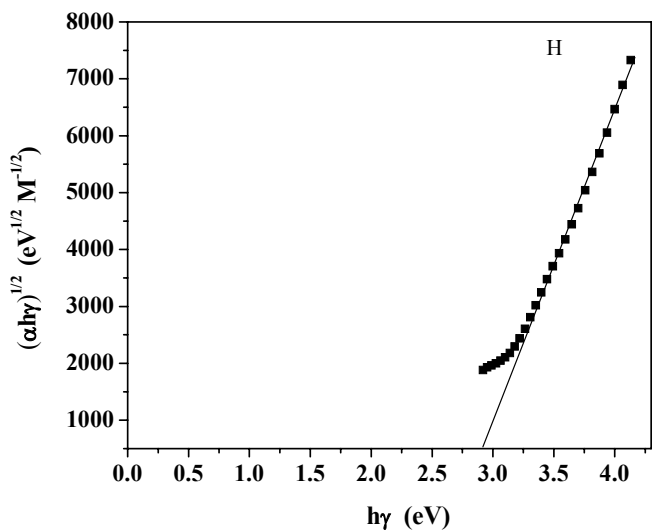

FIG. 4. The $(\alpha h\gamma)^{1/2}$ verses $h\gamma$ of cadmium stannate film

The average value of optical transmittance in the visible region and the band gap energy values obtained by the early investigators are compared with the corresponding values of this work in Table III.

TABLE III. Comparison of optical transmittance in the visible region and band gap energy value.

| S. No | Average Optical Transmittance (%) | Band Gap Energy (eV) | Reference |
|---|---|---|---|
| 1. | 85 | 2, 2.9[a] | 12 |
| 2. | 85 | 2, 2.9[a] | 13 |
| 3. | 85 | - | 18 |
| 4. | 90 | 2.4, 2.7[a] | 19 |
| 5. | 85 | 2.1, 2.9[a] | 7 |
| 6. | - | 2.23, 2.53[a] | 15 |
| 7. | 82.5 | - | 25 |
| 8. | 85 | 2.39, 2.45, 2.71 | 31 |
| 9. | 90 | 2.71 | 20 |
| 10. | 85 | 2.55 | 11 |
| 11. | 80 | - | 32 |
| 12. | 80 | 2.58[a] | 23 |
| 13. | - | 2.2 | 28 |
| 14. | 85 | - | 2 |
| 15. | 88 | 2.9 | Present work (G, H) |
| 16. | 82 | 2.9 | Present work (I) |

[a]Post annealed films.

One can notice from this table that the band gap energy value is increased only after post annealing the films in the Ar-CdS atmosphere at 600-700$^0$ C,[12,13] and Ar-CdS atmosphere at 750 $^0$ C,[7] known as Burstein shift,[1,7,12,13,15,19,23] to a maximum of 2.9 eV.

Thus it is evident that the films prepared in this work without post annealing them in the atmosphere of any such gases give a band gap value of about 2.9 eV.

## IV CONCLUSIONS

Transparent conducting oxide of cadmium stannate films were prepared by employing home made spray pyrolysis technique using low cost precursors for the first time. Film prepared at the substrate temperature of $450^0$ C is amorphous and films prepared at $500^0$ C have the growth orientation along (111) in orthorhombic structure. Sheet resistance of about 160 ohms/square is obtained for the films prepared without additional layer. The process of formation of additional layer decreases the sheet resistance to 15 ohms/square. The maximum optical transmittance of cadmium stannate film of this work is 99.8%. Further all the films prepared in this work have a band gap energy value of about 2.9 eV without post heat treatment in the atmosphere of Ar, CdS, Ar-CdS, hydrogen and nitrogen. As the process of formation of additional layer improves the electrical properties of this film without affecting its structural and optical properties, the formation of additional layer seems to be an effective alternate method for post annealing treatment followed in the other methods to prepare cadmium stannate films.

## ACKNOWLEDGMENTS

The authors gratefully acknowledge Prof S. A. Shiva shankar, Materials Research Centre, Indian Institute of Science, Bangalore and Dr. M. Krishnan, Department of Eco-Biotechnology, Bharathidasan University, Tiruchirappalli – 24 for extending the facility of their laboratory. One of the authors (V. K. K) acknowledges the Ministry of Non-conventional Energy Sources, Government of India for the award of NRE Junior Research Fellowship.

* Corresponding author. Electronic address: krmurthin@yahoo.co.in